\documentclass[12pt]{article}
\usepackage{epsf,epsfig,graphicx}

\newcommand{\sla}{\kern -5.4pt /}
\newcommand{\Dir}{\kern -6.4pt\Big{/}}
\newcommand{\Dirin}{\kern -10.4pt\Big{/}\kern 4.4pt}
\newcommand{\DDir}{\kern -7.6pt\Big{/}}
\newcommand{\DGir}{\kern -6.0pt\Big{/}}

\newcommand{\be}{\begin{equation}}
\newcommand{\ee}{\end{equation}}
\newcommand{\bea}{\begin{eqnarray}}
\newcommand{\eea}{\end{eqnarray}}
\newcommand{\beanon}{\begin{eqnarray*}}
\newcommand{\eeanon}{\end{eqnarray*}}
\newcommand{\ba}{\begin{array}}
\newcommand{\ea}{\end{array}}
\newcommand{\bd}{\begin{description}}
\newcommand{\ed}{\end{description}}
\newcommand{\bi}{\begin{itemize}}
\newcommand{\ei}{\end{itemize}}
\newcommand{\ben}{\begin{enumerate}}
\newcommand{\een}{\end{enumerate}}
\newcommand{\bc}{\begin{center}}
\newcommand{\ec}{\end{center}}

\newcommand{\bqas}{\begin{eqnarray*}}
\newcommand{\eqas}{\end{eqnarray*}}

\newcommand{\ar}{\rightarrow}
\newcommand{\parno}{\par\noindent}

\newcommand{\hsk}{\hskip 10 pt\noindent}

\newcommand{\phact}{{\tt PHACT }}

\newcommand{\phase}{{\tt PHASE }}
\newcommand{\phasenosp}{{\tt PHASE}}

\def\pl #1 #2 #3 {{\it Phys.~Lett.} {\bf#1} (#2) #3}   
\def\np #1 #2 #3 {{\it Nucl.~Phys.} {\bf#1} (#2) #3}
\def\zp #1 #2 #3 {{\it Z.~Phys.} {\bf#1} (#2) #3}
\def\pr #1 #2 #3 {{\it Phys.~Rev.} {\bf#1} (#2) #3}
\def\prep #1 #2 #3 {{\it Phys.~Rep.} {\bf#1} (#2) #3}
\def\prl #1 #2 #3 {{\it Phys.~Rev.~Lett.} {\bf#1} (#2) #3}
\def\intj #1 #2 #3 {{\it Int. J. Mod. Phys.} {\bf#1} (#2) #3}
\def\mpl #1 #2 #3 {{\it Mod.~Phys.~Lett.} {\bf#1} (#2) #3}
\def\rmp #1 #2 #3 {{\it Rev. Mod. Phys.} {\bf#1} (#2) #3}
\def\cpc #1 #2 #3 {{\it Comp. Phys. Commun.} {\bf#1} (#2) #3}
\def\epj #1 #2 #3 {{\it Eur. Phys. J.} {\bf#1} (#2) #3}
\def\xx #1 #2 #3 {{\bf#1}, (#2) #3}

\begin{document}
\tolerance=100000



{\hfill DFTT 14/2005}

\begin{center}
{\Large \bf Boson boson scattering at LHC}
 \\

\vspace{4mm}

  Elena Accomando, Alessandro Ballestrero, Aissa Belhouari and 
    Ezio Maina \footnote{To appear in Proceedings of the XVIII Workshop on
Quantum Field Theory and High Energy Physics (QFTHEP04) - St. Petersburg
 June 2004. \\
 Work supported by the European Union (EU) under contract HPRN-CT-2000-00149
and by MIUR under contract 2004021808\_009. . 
The work of E.A. is supported by the MIUR under Contract 
``Rientro dei Cervelli'' Decreto MIUR 26-01-2001 N.~13}
\\
INFN, Sezione di Torino and  Dipartimento di Fisica Teorica, Universit\`a di
Torino\\
Via P. Giuria 1, 10125, Torino, Italy\\
\end{center}

\begin{abstract}
We  analyse some features of WW scattering processes at LHC. 
The severe cancellations between fusion diagrams and the other
contributions evidence the necessity of complete calculations
for  studying   the high WW invariant mass region and disentangling
the standard Higgs case from new physics.
\end{abstract}

\section{ Introduction }
 Higgs discovery and Electroweak Symmetry Breaking (EWSB) studies are among the  main purposes of LHC. The so called boson boson fusion  
process is one of the most promising Higgs production channels. If the Higgs mass
is greater than 140 GeV, its main decay mode is in two W's.
The study of this channel requires a complete calculation 
of six fermion final states.   On the other hand it is well known that
a no-Higgs scenario would imply a violation at high energies of partial
wave perturbative unitarity, which would manifest itself in particular
in WW scattering processes. In this sense WW scattering processes hold
the key of EWSB. Possible new physics, which should account for the
restoration of unitarity, would manifest itself there.
As it is experimentally impossible to measure WW scattering,
there is great interest in studying at LHC processes of the
type $PP \ar q q' V V $ ($V=W,Z$), which contain boson boson fusion
diagrams. One expects that this contribution dominates in particular 
kinematical regions and that it is possible from these studies
to gain a better understanding of EWSB.

In the following we consider some theoretical difficulties connected with
this kind of physics and we argue that it is necessary
a  full control of all six fermion processes.
Some of the results presented are obtained with \phasenosp \cite{phase} , a 
new Monte Carlo and event generator aimed at this scope.

\section{ From six fermion final states to WW scattering}

In order to study WW scattering effects, one has to explore six fermion 
final states at LHC.
\begin{figure}[htb]
  \begin{center}
  \unitlength 1cm
 \begin{picture}(15,3.5) (0,0)
  \put(-.5,0.){ \scalebox{.35}{\includegraphics*{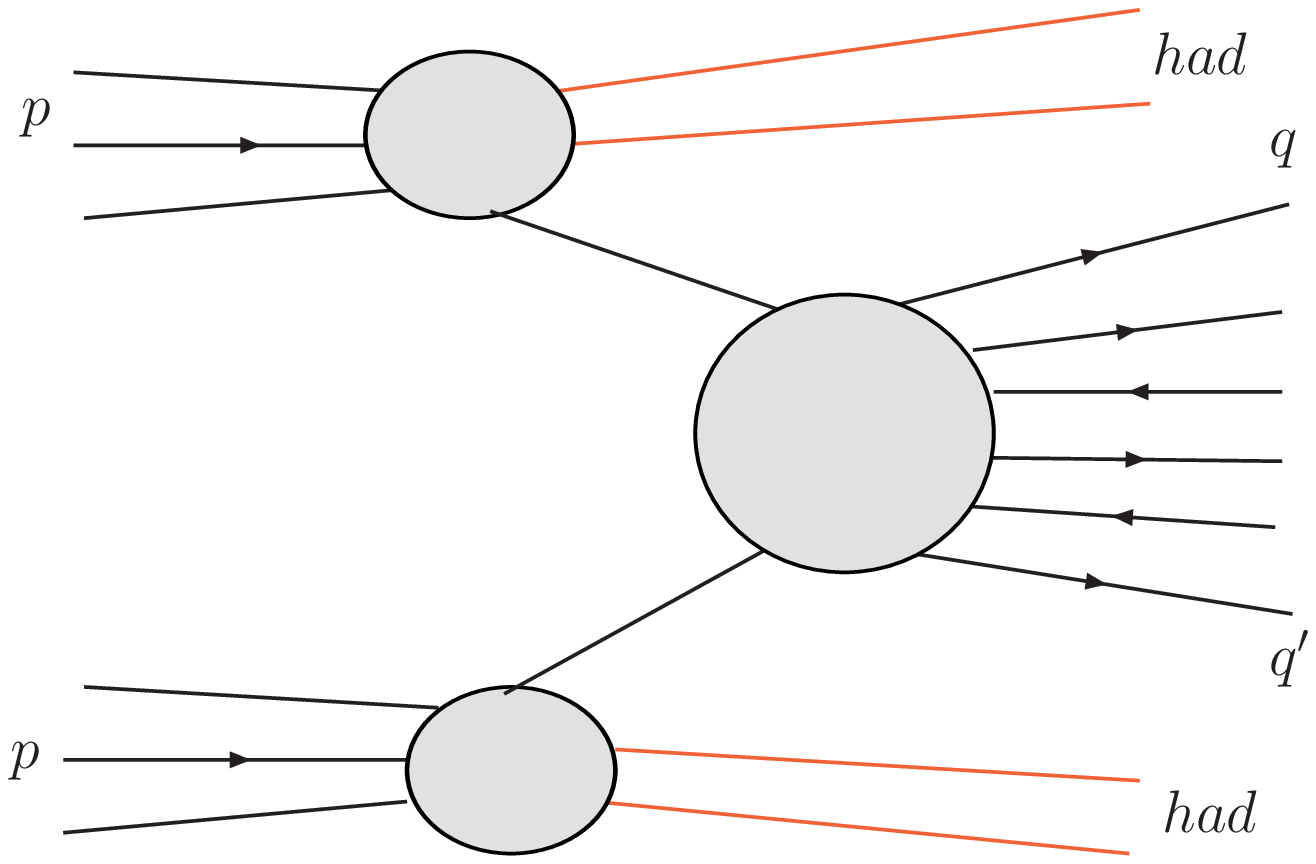}} }
  \put(5.,0.){ \scalebox{.35}{\includegraphics*{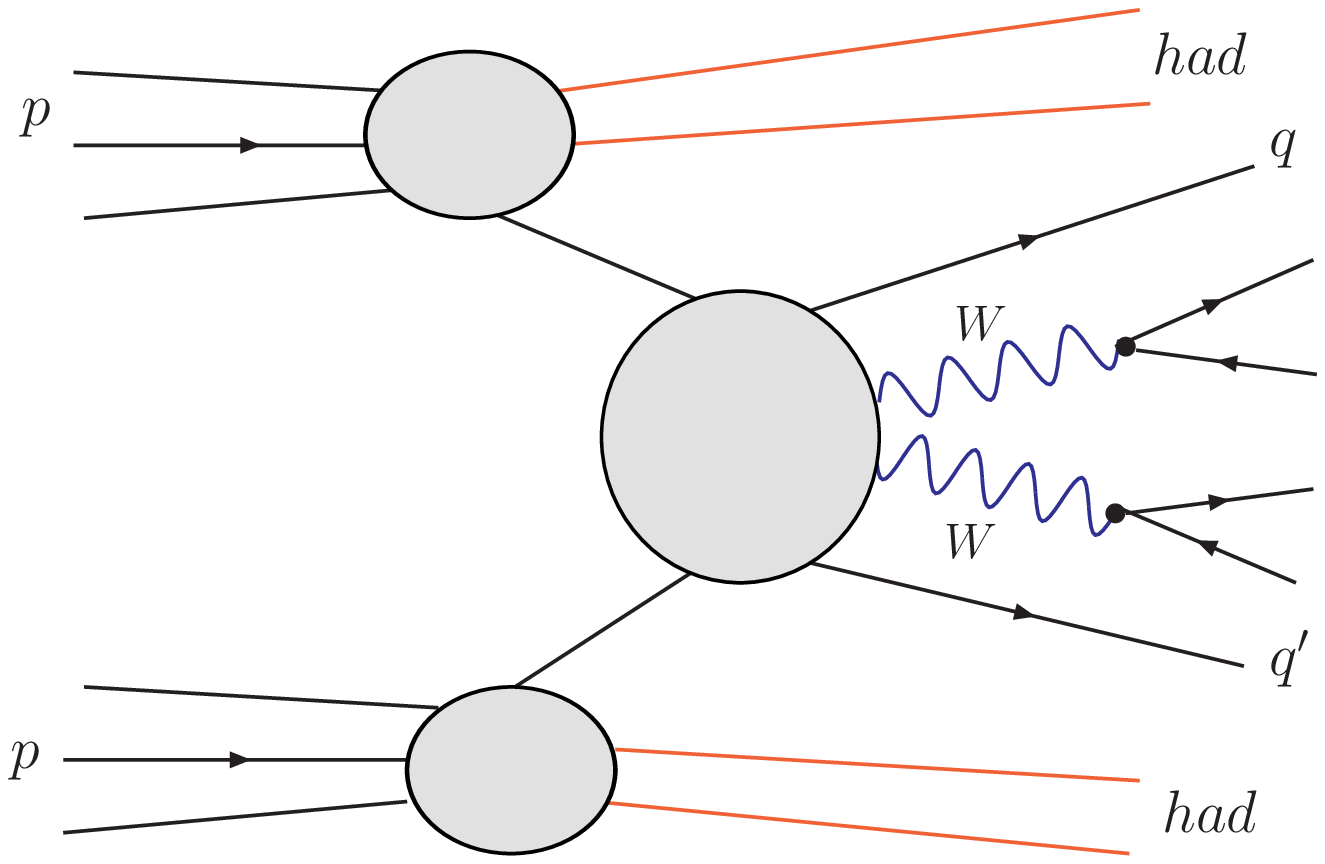}} }
  \put(10.5,0.){ \scalebox{.35}{\includegraphics*{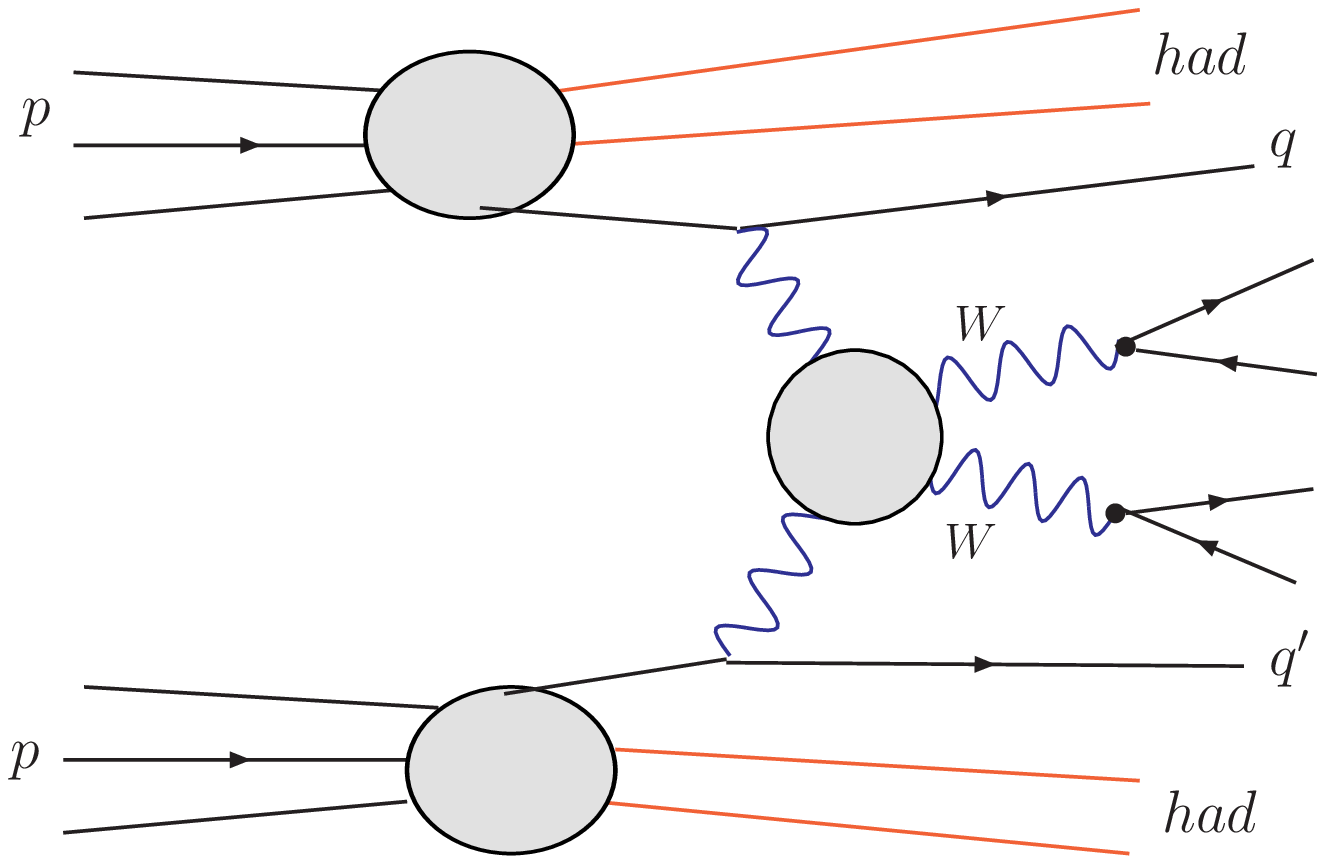}} }
  \put(-.4,1.25) {a)}
  \put(5.2,1.25) {b)}
  \put(10.6,1.25) {c)}
\end{picture}
\caption[]{ Representation  of six fermion processes at LHC (a),
$qq'WW$ production (b) and boson boson scattering  contributions (c).  
}
\ec
\end{figure}
The situation is depicted schematically in Fig.~1. 
Six fermion processes  initiated by two partons (a),
 contain   contributions from a set
of diagrams with two decaying W's (b). These in turn contain diagrams with
two off shell bosons ``emitted'' by the incoming quarks which scatter
to the final state W's (c). These latter contributions are normally indicated
as boson boson fusion.

Ideally,  informations on WW scattering should be deduced isolating  
the contributions of the diagrams of Fig.~1 (c), deconvoluting them from 
the pdf's of the incoming partons and extrapolating the WW scattering 
subdiagrams to on shell incoming W's. 
This is a formidable task, and we will consider in the following some
theoretical reasons why we believe that it cannot be accomplished as such.
It seems  however possible to find relevant differences between the
Higgs and no-Higgs case, especially in distributions of the invariant
mass of the two final W's, which is the variable that in this context 
plays the role of the center of mass energy in the pure
WW scattering process.

\section{ WW scattering and gauge invariance}

Let us now consider the hard process $qq'\ar q q' W^+W^-$.
The corresponding Feynman diagrams can be classified in different topologies,
as shown in Fig.~2. 

\begin{figure}[htb]
  \begin{center}
  \unitlength 1cm
 \begin{picture}(15,3.) (0,0)
  \put(0.,0.){ \scalebox{.35}{\includegraphics*{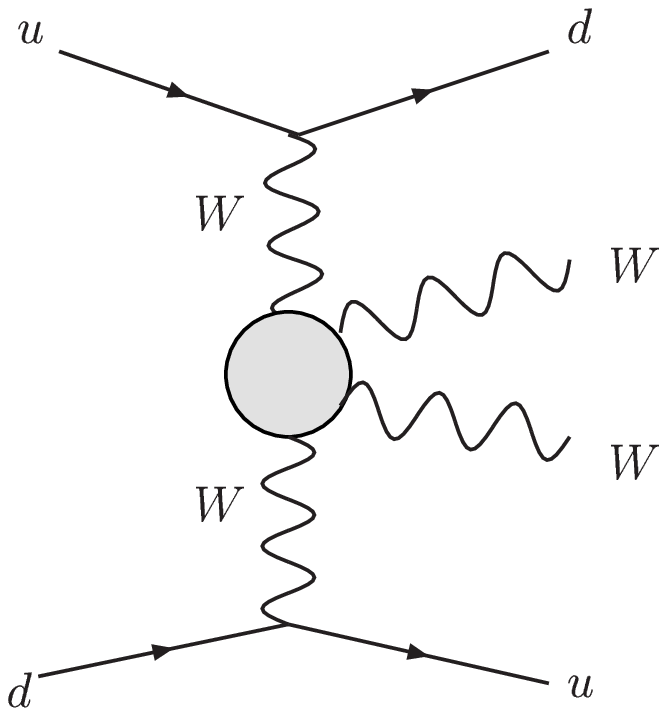}} }
  \put(4.,0.){ \scalebox{.35}{\includegraphics*{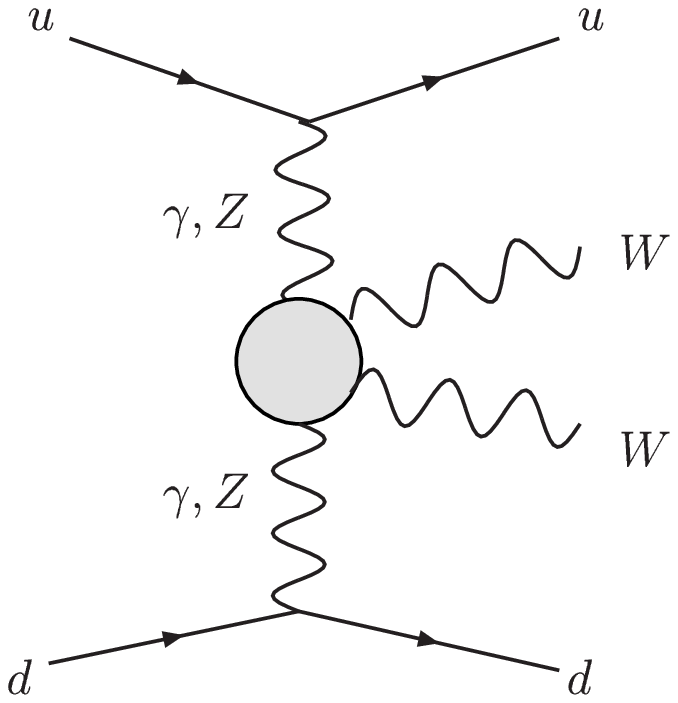}} }
  \put(8.,0.){ \scalebox{.35}{\includegraphics*{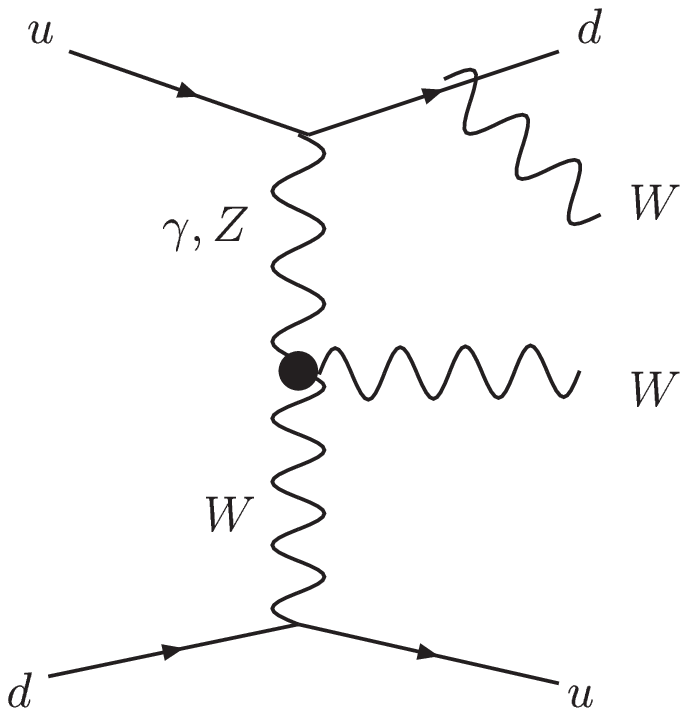}} }
  \put(12.,0.){ \scalebox{.35}{\includegraphics*{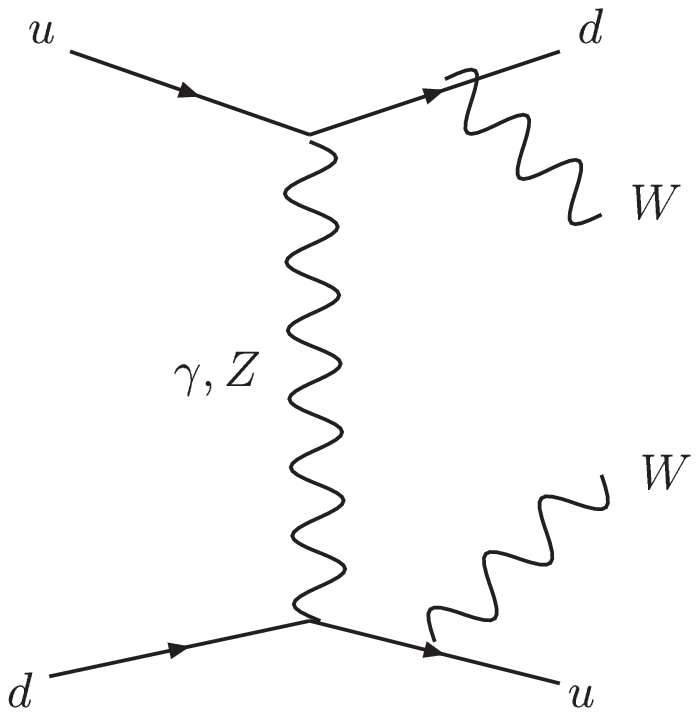}} }
  
\end{picture}
\caption[]{Different types of Feynman diagrams for the process  
$qq'\ar q q' W^+W^-$
}
\ec
\end{figure}

Essentially one has two contributions: boson boson fusion 
(first two on the left) and all the rest in which at least
one of the final W's is emitted by fermion lines. Only the former belongs
 to Fig.~1 (c). The two sets are not separately gauge invariant and considering
only one of them may be dangerous.

It is  known since a long time\cite{KS} that there are
large cancellations from negative interference between boson boson fusion 
diagrams and the rest. We have analysed in detail such effects for the sample
process  $PP\ar us \ar dc W^+W^-$, in order to understand if one can find
a particular gauge and a  kinematical region in which the cancellations at LHC 
are not so severe. Some of our results are reported in Table~1 and in 
Figs~3,4.

\begin{table}[htb]

\bc
\begin {tabular}{||c|c|c ||}  \hline
      no-Higgs            &  $\sigma$ (pb)  &   ratio   WW/all  \\  \hline
all diagrams              &  1.86 E-2       &                   \\  \hline
WW fusion (unitary gauge) &  6.67           &    358            \\  \hline
WW fusion (Feynman gauge) &  0.245          &    13             \\  \hline
\end {tabular}
\ec
\caption[] {Cross sections for the process $PP\ar us \ar dc W^+W^-$ evaluated
using all diagrams or only the WW fusion subset, and  their ratio. }
\end{table}

We find that, among the various $R_\xi$ gauges,  the Feynman gauge is the
one which diminishes the interference. However, we see from Table~1 that 
even in this case the separate contributions of WW fusion and of the remaining
diagrams have total cross sections an order of magnitude larger than the 
exact one, computed with all diagrams. The numbers reported in Table~1
refer to the no-Higgs case. For $M_H$=200 GeV in the
region of invariant masses above the Higgs peak ($M_{WW}>300$ GeV),
the ratios are even larger, by about a factor of 2.   

\begin{figure}[htb]
  \begin{center}
  \unitlength 1cm
 \begin{picture}(15,4.5) (0,0)
  \put(.5,0.){ \scalebox{.5}{\includegraphics*{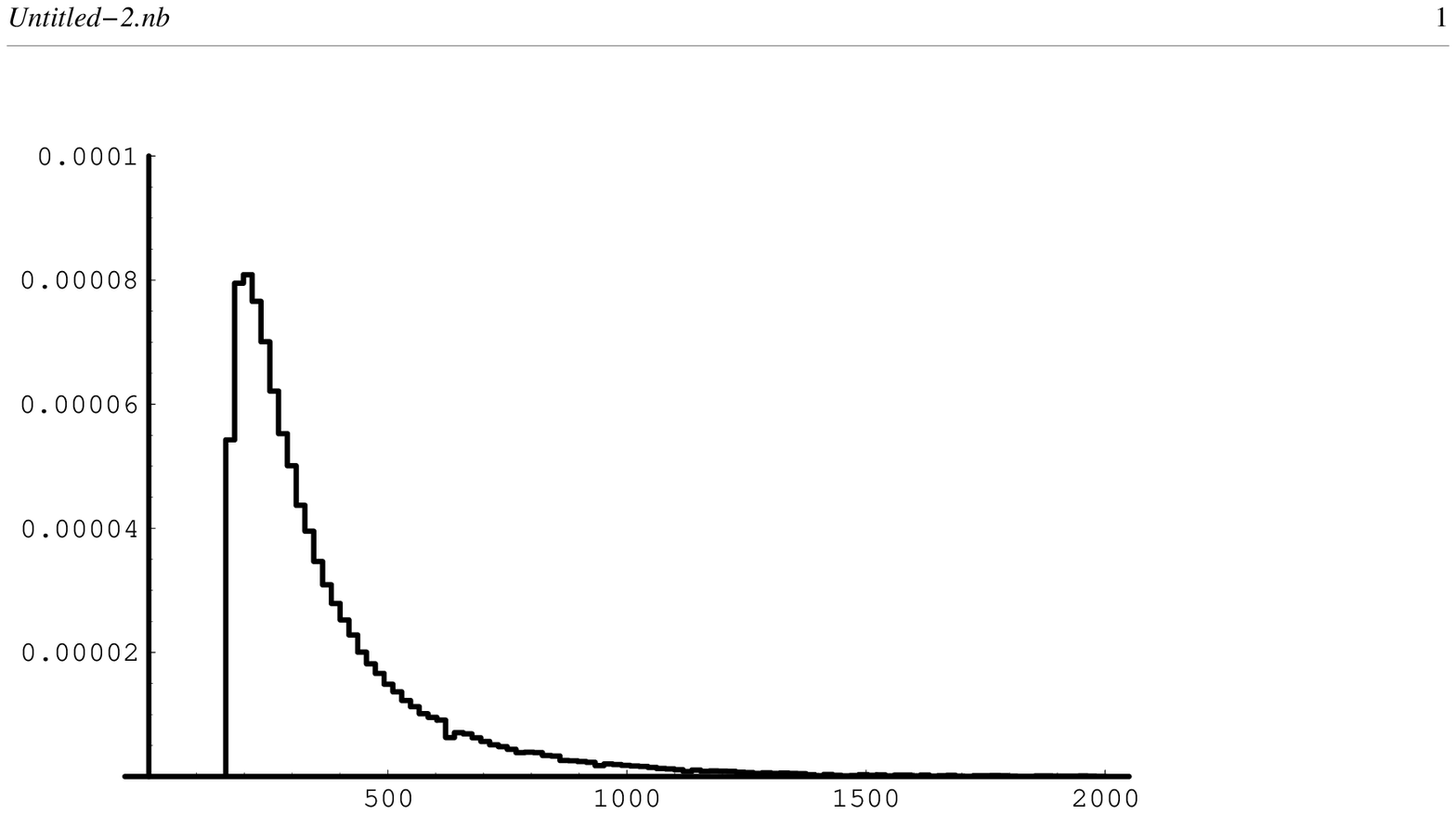}} }
  \put(7.4,0.){ \scalebox{.5}{\includegraphics*{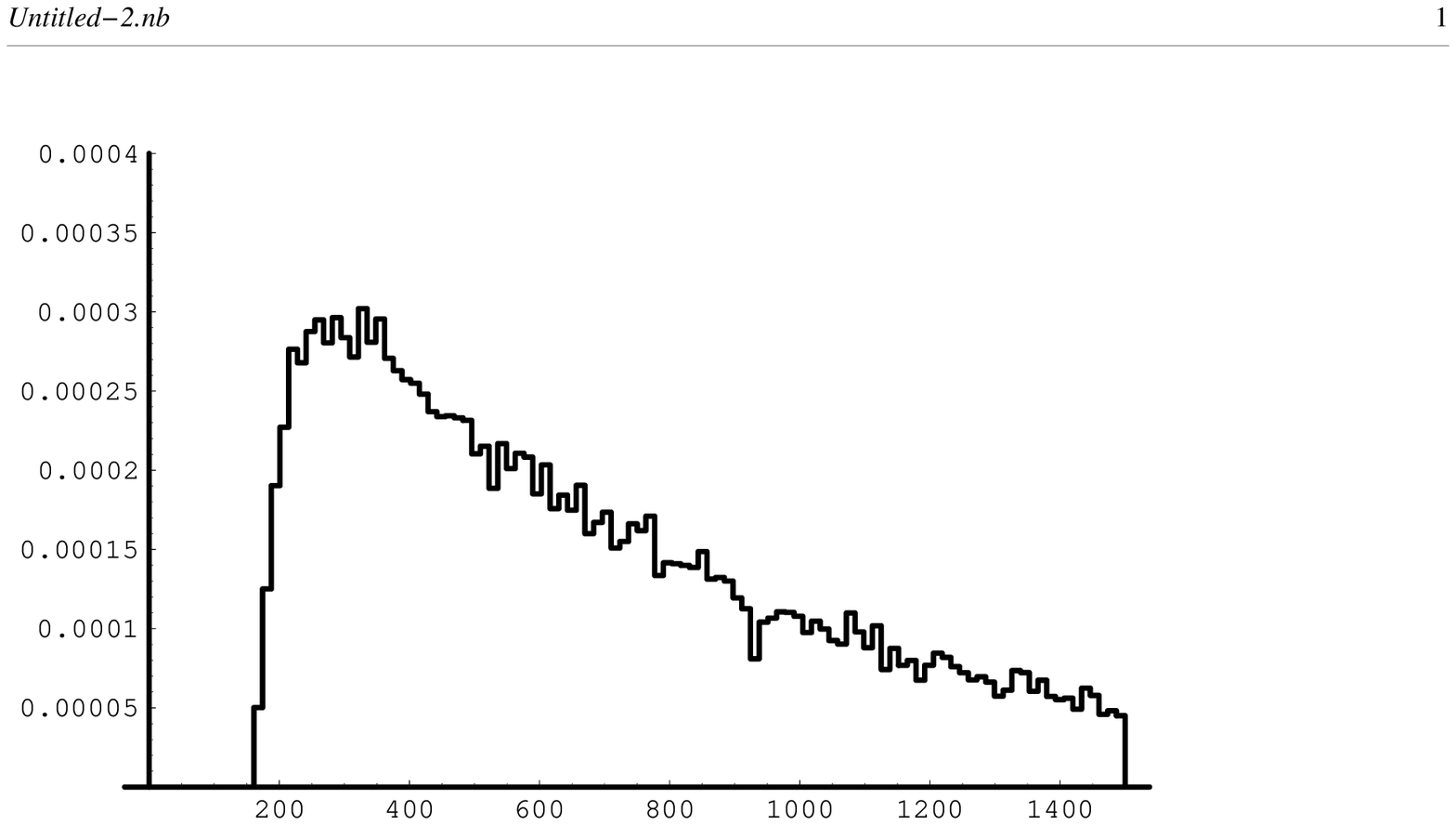}} }
  \put(.1,2.5) {a)}  
  \put(7.,2.5) {b)}  
\end{picture}
\caption[]{Distributions  of $d\sigma/dM_{WW}$ $(pb/GeV) $ 
for the process $PP\ar us \ar dc W^+W^-$,
using all diagrams 
(a) or only the WW fusion subset in Feynman gauge (b).
}
\ec
\end{figure}

Considering now the distributions reported in Fig.~3, one realizes that
the shape of the curve for total (left) and WW fusion only (right)  is 
quite different and that the ratio between the two curves grows
with the WW invariant mass. This effect is much more relevant in unitary gauge.

\begin{figure}[htb]
  \begin{center}
  \unitlength 1cm
 \begin{picture}(15,11) (0,0)
  \put(.7,5.8){ \scalebox{.45}{\includegraphics*{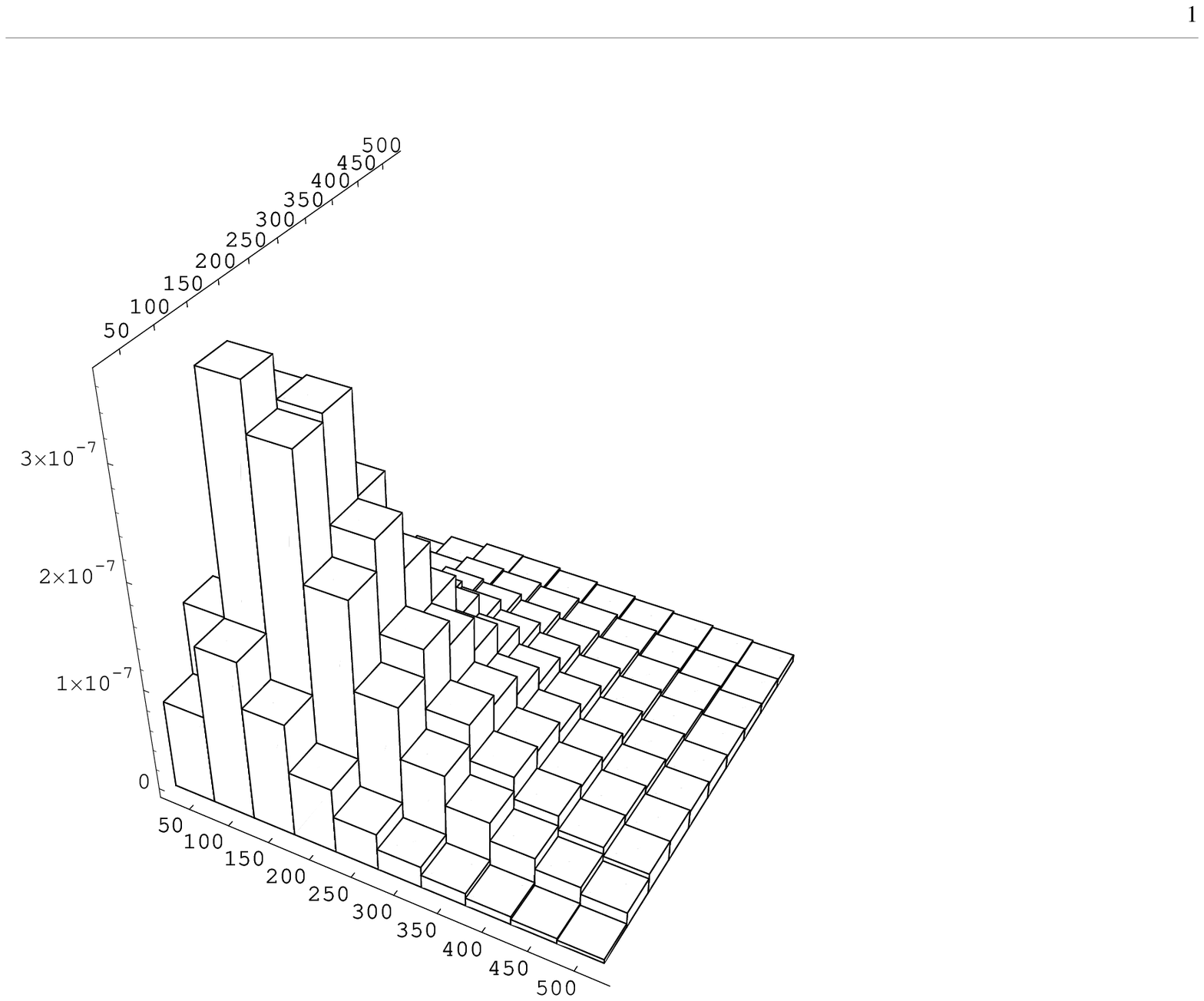}} }
  \put(8.,5.8){ \scalebox{.45}{\includegraphics*{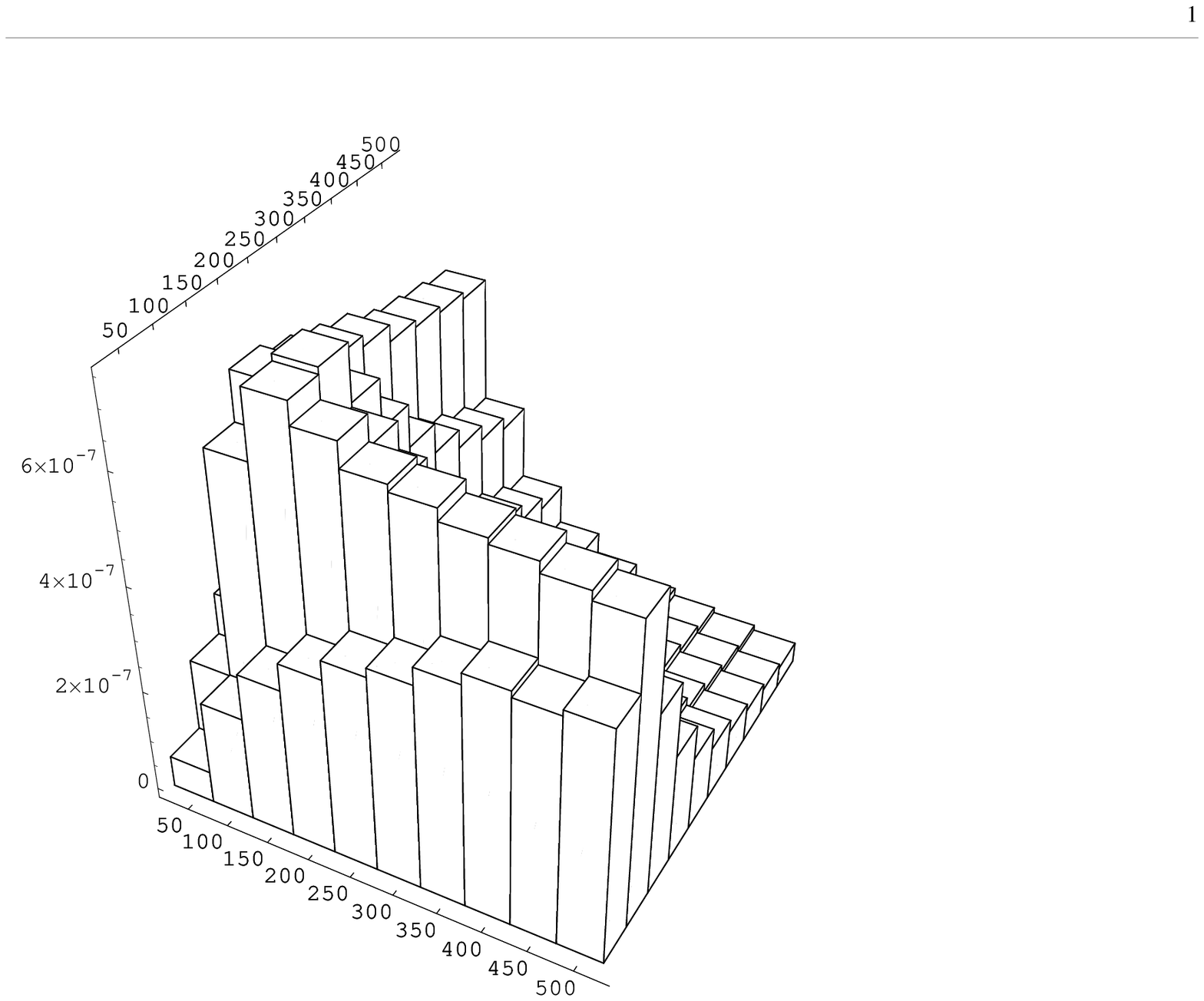}} }
  
  \put(.7,.3){ \scalebox{.4}{\includegraphics*{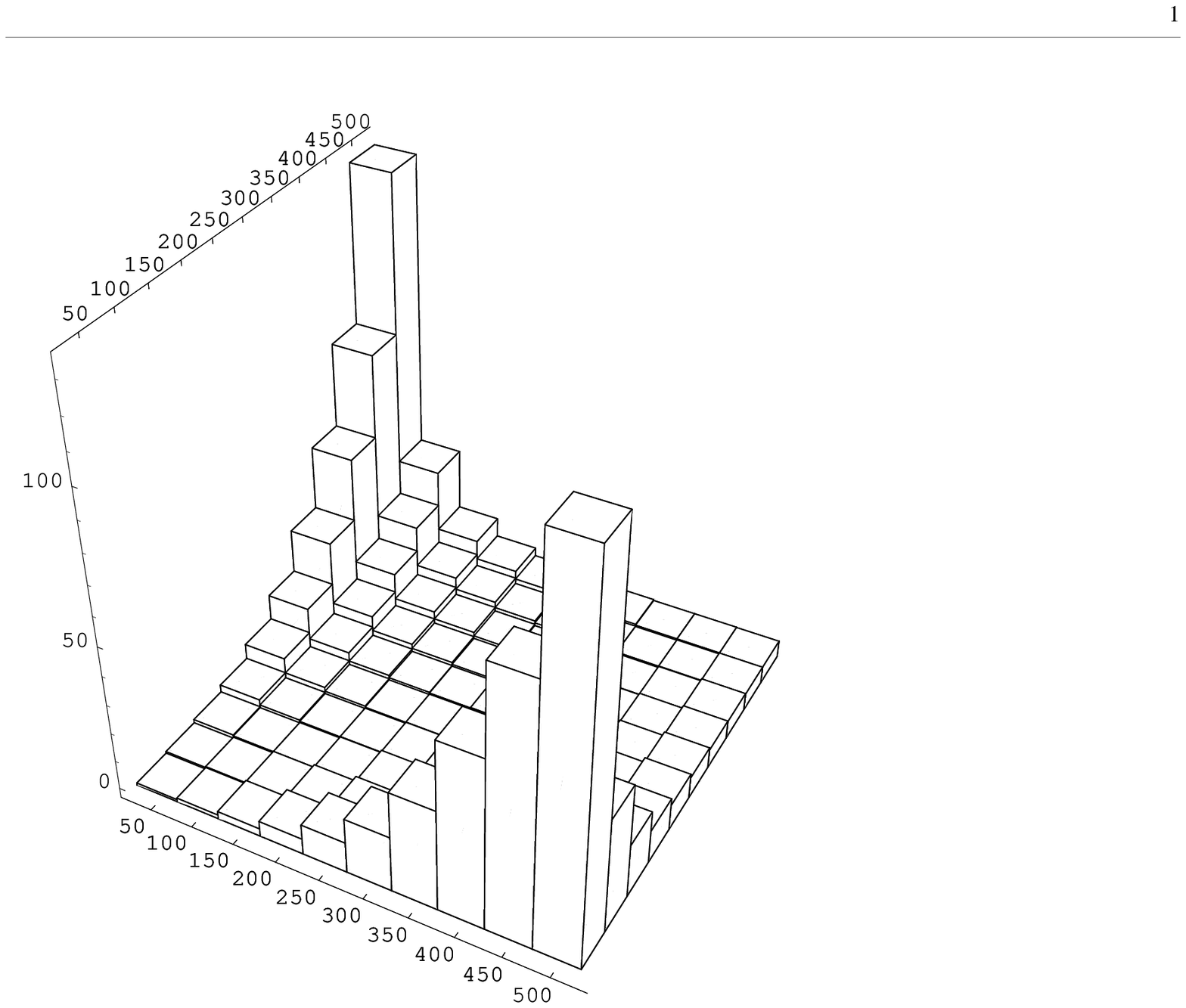}} }
  \put(8.,0.){ \scalebox{.4}{\includegraphics*{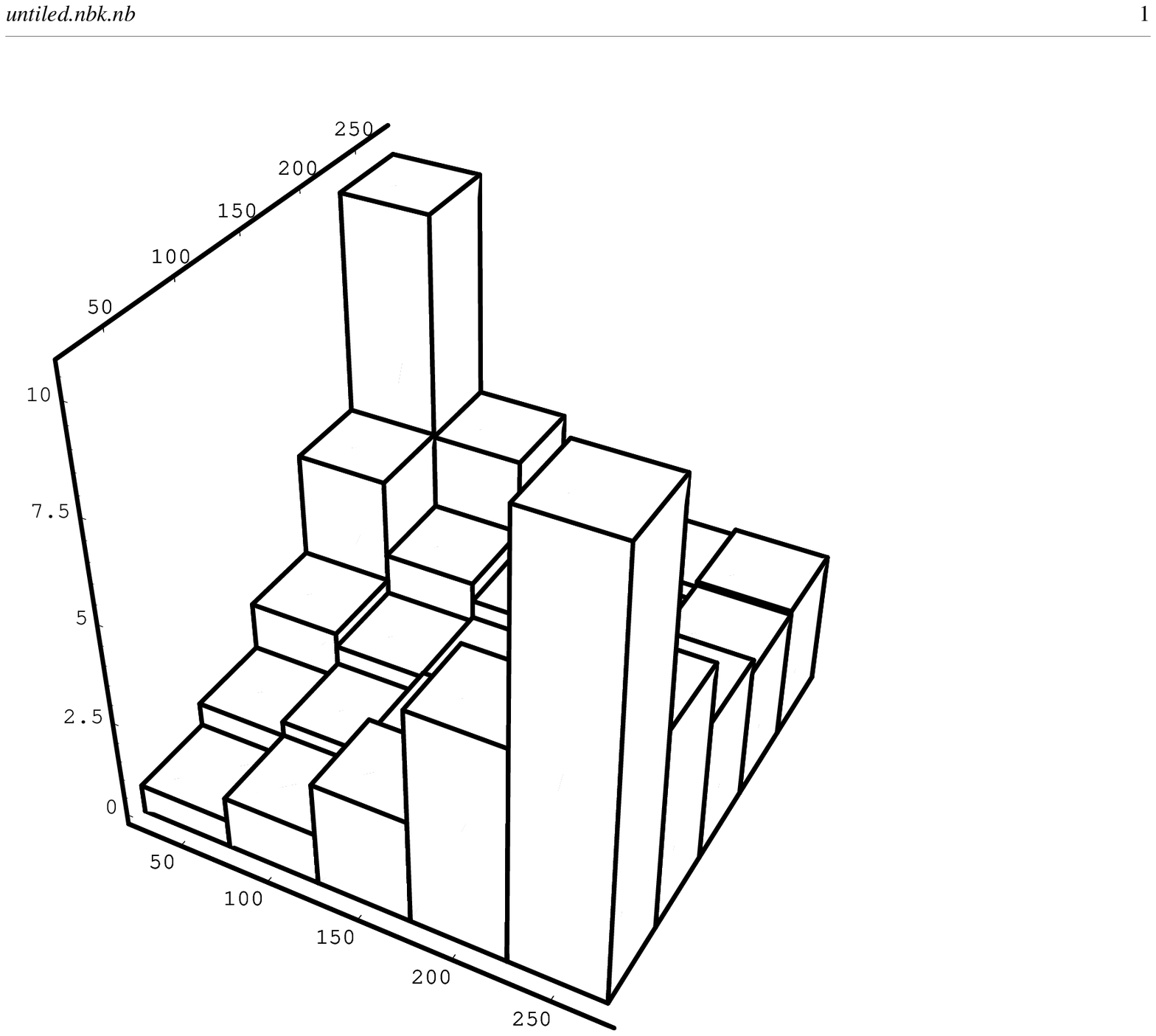}} }

  \put(.2,8.5) {a)}  
  \put(7.4,8.5) {b)}  
  \put(.1,2.5) {c)}  
  \put(7.3,2.5) {d)}  
 
\end{picture}
\caption[]{Double differential distributions of $d\sigma/dt_1dt_2$ 
$(pb/GeV^2) $  with $t_{1,2}=\sqrt{-(p_{u,s}-p_{d,c})^2}$ for the process 
$PP\ar us \ar dc W^+W^-$ , using all diagrams 
(a) or only the WW fusion subset in Feynman gauge (b).
The remaining plots represent the ratio WW/all as a function of $t_1,t_2$ 
for the total region (c) and for small $t$'s (d).
}
\ec
\end{figure}

Fusion diagrams are expected to  peak in the region
of low momentum transfer from incoming quarks. Therefore we
have analysed in Fig.~4 the double differential distributions 
$d\sigma/dt_1dt_2$  with $t_{1,2}=\sqrt{-(p_{u,s}-p_{d,c})^2}$ 
for the process $PP\ar us \ar dc W^+W^-$. Again the conclusion is
that the   WW fusion subset has a different behaviour compared with the
complete calculation. The interference effects are non negligible 
even for moderate values of  $t_{1,2}$ where
the complete cross section is larger. 

The plots of Fig.~4 have been obtained excluding contributions from Higgs 
diagrams ($m_H \ar \infty$ limit). The
qualitative behaviour does not change if one includes them and
excludes  the region $M_{WW} \approx M_H$.

 The Equivalent Vector Boson Approximation (EVBA) \cite{evba}, which
is gauge independent, is often used to estimate the contribution of 
the boson boson fusion. The validity of such an approximation
for massive vector bosons has been studied in the literature \cite{evbares}. 
It appears that at LHC energies the EVBA is a reasonable approximation 
to the complete cross section for Higgs production but not for  boson boson 
invariant masses away from the peak.
Moreover distributions such as those in fig~4~(a) cannot be correctly 
reproduced. 

\section{ \phase Monte Carlo}

The strong cancellations examined in the previous section indicate that it is
necessary to use complete calculations for the analysis of boson boson 
physics. We believe that also considering qqWW final state and then
decaying the W's to 4 fermions is an approximation to the complete six fermion
calculations that might not be sufficient for some analyses. We will give an
example of this in the next section.

The first version of the new Monte Carlo and event generator \phase 
\cite{phase}  for six fermion final states   has  recently been completed.
This is a Monte Carlo for LHC dedicated studies with full physics and detector
simulation of 
 boson boson fusion and scattering,
Higgs production in this channel,
$t\bar t$ production,
triple and quadruple boson couplings,
three boson production.

The first version of \phase can be used for processes with 4 quarks, 
a lepton and a neutrino in the final state. Processes are computed at 
{\sl O}($\alpha_{em}^6$). 

Matrix elements are evaluated with routines written with \phact\cite{phact}.
The number of different processes that can at present be computed is of the
order of one thousand, each described by hundreds Feynman diagrams.
The calculation is organized in such a way that only two groups of 
respectively 101 and 305 independent diagrams are recursively evaluated with
a technique which makes repetitive use of subdiagrams to get a faster
evaluation. \phase adopts a new integration method which combines 
adaptive and multichannel strategies. High precision
and efficiency has been obtained. \phase can be used to evaluate any individual
process or in the so called one-shot mode for generation\cite{wphact}. In 
the latter, events of all processes or any subset
can be generated at once in the right proportion, after the integrand maxima
and phase space grids for the various processes have been computed 
in a preparatory run.
 
\section{ Boson Boson fusion and Higgs at high WW masses}

In order to study in more detail 
 the consequences of neglecting part of the calculations and to find out
whether also in the complete calculations one can evidentiate differences
between Higgs and no-Higgs case, we examine in this section
the process $PP\ar ud  \ar ud c\bar s \mu \bar \nu_\mu $.
\begin{figure}[htb]
  \begin{center}
  \unitlength 1cm
 \begin{picture}(15,7) (0,0)
 \put(1.5,7.){ \scalebox{.45}{\includegraphics*[angle=-90]{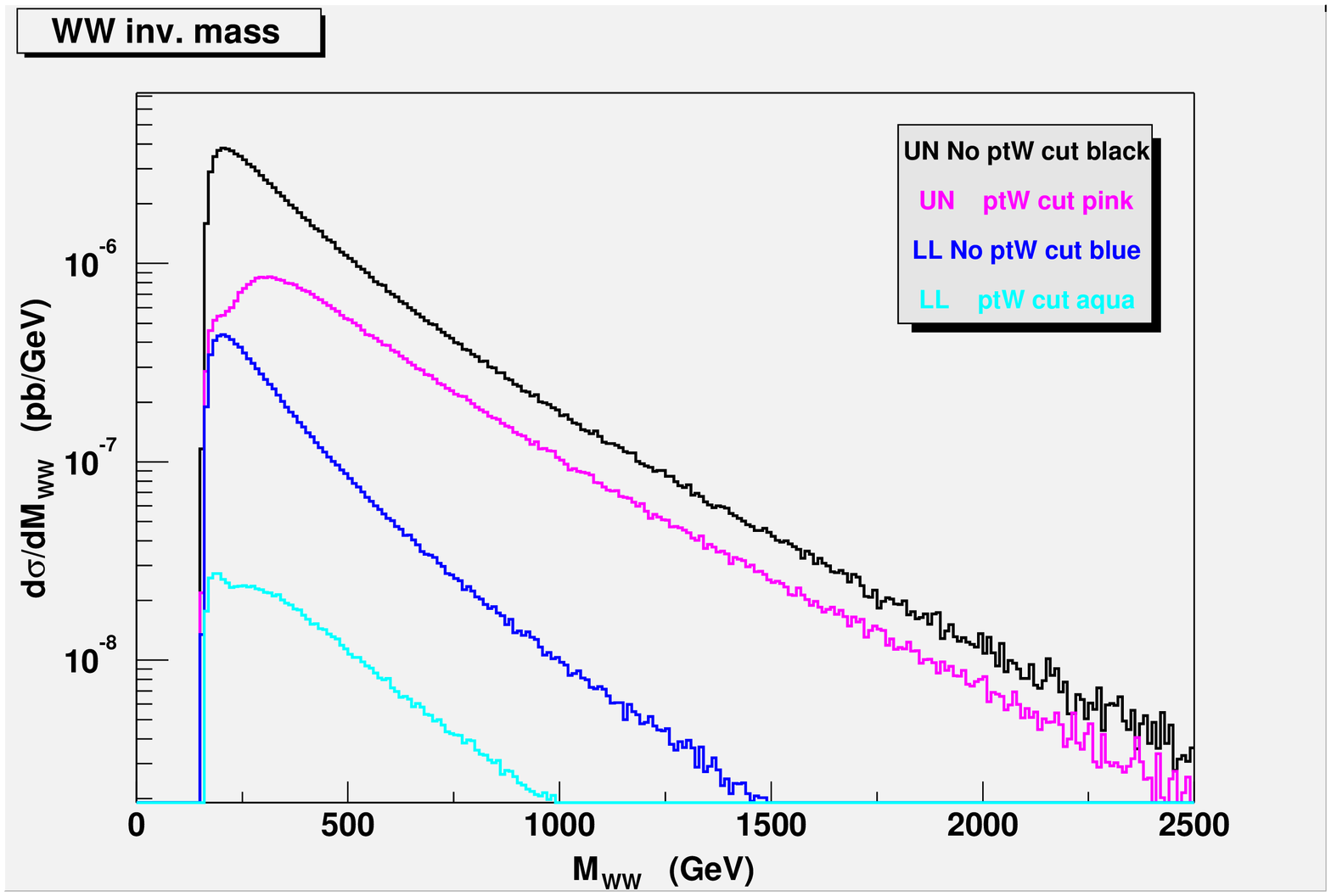}} }
  
\end{picture}
\caption[]{Differential distributions for the process 
$PP\ar ud  \ar ud W^+W^- \ar ud c\bar s \mu \bar \nu_\mu $ with 
$m_H =120$ GeV for unpolarized and longitudinally polarized W's. 
Cuts are described in the text
}
\ec
\end{figure}
One can select the resonant diagrams  
$PP\ar ud  \ar ud W^+W^- \ar ud c\bar s \mu \bar \nu_\mu $  and analyse
the $M_{WW}$ distributions. Moreover, even if the $W$'s  
are off shell, one can define and separate Longitudinal and 
Transverse final polarizations. 
We study in fig.~5 the  $M_{WW}$ distributions for the Unpolarized case
(UN) and Longitudinal Longitudinal (LL) one. 
It is expected that EWSB effects manifest themselves in this distribution
and that they are more pronounced for the LL case. 
We have imposed the following cuts:

$1<\eta (d)<5.5$ \hsk $-1>\eta (u)>-5.5$ \hsk $70 < M(sc,\mu \nu) < 90$ GeV 

$E(u,d,c,s,\mu) > 20$ GeV \hsk $P_T(u,d,c,s,\mu) > 10$ GeV.
\parno
It is evident, from the first and third curve of fig.~5 from above, 
that considering only LL contributions underestimates the differential cross 
section by at least a factor ten. The exact ratio is strongly dependent
on the Higgs mass.

In the other two curves we have required  $p_T^W > M_W$. 
A cut of this kind is unavoidable if one uses the 
EVBA approximation, in order to keep under control the divergence of  
the on shell $WW\ar WW$ diagram with $\gamma$ exchange in t channel. 
One can see from the second and fourth curve from the top that in this case 
the difference between LL and UN is even larger.
 
\begin{figure}[htb]
  \begin{center}
  \unitlength 1cm
 \begin{picture}(15,8.) (0,0)
 \put(1.5,8.){ \scalebox{.45}{\includegraphics*[angle=-90]{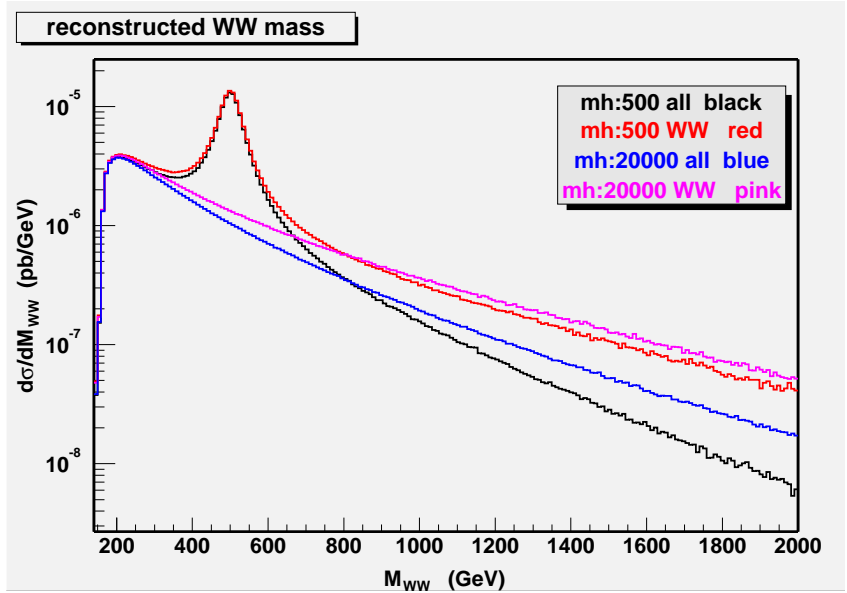}} }
  
\end{picture}
\caption[ ]{Reconstructed WW mass distributions for the process
 $PP\ar ud  \ar ud c\bar s \mu \bar \nu_\mu $ 
in the Higgs and no-Higss 
($m_H \ar \infty$) case, with the full set and WW resonant only diagrams.
Cuts are defined in the text.
}
\ec
\end{figure}

In fig.~6 we study the differences between Higgs and no-Higgs 
($m_H \ar \infty$) scenarios for the full six fermion calculation and for the 
case in which only resonant 
$ PP\ar ud  \ar ud W^+W^- \ar ud c\bar s \mu \bar \nu_\mu $ diagrams
are used.
  
We plot the ``recostructed'' WW mass. By this we mean the following:
we select the most forward (f) and the most backward (b) jet, and
call the remaining jets central (c).
The invariant mass of the two central jets, the lepton and the neutrino, 
is what we call ``reconstructed mass''.
We impose the following cuts:

$2<\eta (j_f)<6.5$ \hsk $-2>\eta (j_b)>-6.5$ \hsk $|\eta(j_c,\mu )|<3$ \hsk 

$E(j,\mu )>20$ GeV \hsk $P_t(j,\mu )>10$ GeV \hsk $M(jj)>20$ GeV \hsk 

$|M(j_c,j_c)-M_W|<20$ GeV \hsk $M_T(\mu \bar \nu_\mu ) <M_W + 20$ GeV,
\parno 
One may notice from fig.~6 that the two upper curves, corresponding to 
only resonant diagrams, substantially differ in the high mass region
from the complete calculation. It is even more important
to outline  that the two lower curves, describing 
the complete calculation for the Higgs and no-Higgs scenario, differ among 
themselves by a factor $\approx 2\div 3$. This is encouraging with respect to
the possibility of finding signals of EWSB at LHC. 
 
\section{ Conclusions}
 
We have found that the extraction of boson boson scattering contributions
at LHC can be problematic due to gauge invariance and cancellation effects.
There are nevertheless clear indications that, if one uses the complete
calculations and relies on appropriate cuts, which still have to be optimized, 
the Higgs and no-Higgs scenarios show appreciable differences. 
If a non SM EWSB  mechanism is at work, much more dramatic phenomena  
would show up in  boson boson invariant mass distributions.
We believe that a realistic study with all processes and full detector 
simulation is worthwhile and needed. \phase 
has been developed to this end.

\section*{Acknowledgments}
A. Ballestrero is grateful to all the organizers of QFTHEP04 
for the pleasant and constructive atmosphere during the workshop and
for the warm hospitality.

\vfill\eject

\end{document}